\documentstyle[aps,eqsecnum]{revtex}

\begin{document}

\draft
\title{\bf Universality of $S$-matrix correlations for deterministic plus 
random Hamiltonians}
\author{N. Mae and S. Iida}
\address{Faculty of Science and Technology, Ryukoku University, Otsu 520-2194, 
Japan}
\date{October 21, 2000}
\maketitle

\begin{abstract}
We study $S$-matrix correlations for random matrix ensembles with a 
Hamiltonian $H=H_{0}+\varphi$, in which $H_{0}$ is a deterministic 
$N \times N$ matrix and $\varphi$ belongs to a Gaussian random matrix 
ensemble. 
Using Efetov's supersymmetry formalism, we show that in the limit 
$N \to \infty$ correlation functions of $S$-matrix elements are universal on 
the scale of the local mean level spacing: 
the dependence of $H_{0}$ enters into these correlation functions only through 
the average $S$-matrix and the average level density. 
This statement applies to each of the three symmetry classes (unitary, 
orthogonal, and symplectic).
\end{abstract}
\pacs{PACS number(s): 05.40.-a, 05.30.-d, 05.60.Gg}

\section{Introduction}

The energy levels and/or the scattering matrices of a variety of physical 
systems with randomness (e.g., complex nuclei, disordered 
conductors, classically chaotic systems, etc.) exhibit universal behavior: 
the statistical properties of the observables can be separated into the 
universal parts and the non-universal parts specific to individual systems. 
There have been increasing evidences that the universal parts depend only on 
the fundamental symmetries of the underlying Hamiltonian and are well 
described by a random matrix ensemble with a Gaussian distribution (see 
Ref.~\cite{GMW} for a review). 
According to the fundamental symmetries, there are three classical random 
matrix ensembles: 
Systems with broken time-reversal symmetry are described by the unitary 
ensemble and time-reversal invariant systems by either the symplectic or the 
orthogonal ensemble depending on whether spin-orbit coupling is present or not 
\cite{Bulk}. 
In spite of their many successful applications, random matrix models lack firm 
foundation. 
Especially, the Gaussian form of the probability distribution is used for 
mathematical convenience and is not motivated by physical principles. 
It is therefore necessary and important to investigate whether statistical 
properties are identical for more general forms of the probability 
distribution consistent with fundamental symmetries. 

There has been several work along this direction. 
Hackenbroich and Weidenm\"uller \cite{HW} considered a non-Gaussian and 
unitary invariant probability distribution: 
$
P(H) \propto \exp \left[ -N \,\mbox{tr}\, V(H) \right]
$, 
where $N$ is the dimension of the Hamiltonian matrix $H$ and $V(H)$ is 
independent of $N$ and arbitrary provided it confines the spectrum to some 
finite interval and generates a smooth mean level density, in the limit 
$N \to \infty$. 
For each of the three symmetry classes, using Efetov's supersymmetry 
formalism, they showed that both energy level correlation functions and 
correlation functions of $S$-matrix elements are independent of $P(H)$ and 
hence universal if the arguments of the correlators are scaled correctly. 

For realistic situations, it is likely that the Hamiltonian is not completely 
random but contains some regular parts. 
If the total Hamiltonian $H = H_{0} + \varphi $ where $H_{0}$ is a 
deterministic part and  $\varphi$ is a random one, the probability 
distribution takes a unitary non-invariant form:
\begin{equation}
P(H) \propto 
\exp\left[ -N \,\mbox{tr}\, V(\varphi) \right] = 
\exp\left[ -N \,\mbox{tr}\, V(H - H_{0}) \right].
\label{eqn:gpdfdpr}
\end{equation}
For the unitary ensemble, Br\'ezin et al. \cite{BHZ} discussed the 
universality of 2-point energy level correlations for 
$V(\varphi) = \varphi^2/2 + g \varphi^4$. 
General n-point energy level correlation functions were shown to be universal 
by Br\'ezin and Hikami \cite{BH} for $V(\varphi) \propto \varphi^2$. 
(The other type of unitary non-invariant distrubution 
$
P(H) \propto 
\exp\left\{ -N \,\mbox{tr}\left[ V(H) - H H_{0} \right]\right\}
$ 
was also considered by Zinn-Justin \cite{Z}.)

Recently, we \cite{MI} numerically found the same universality of the 
$S$-matrix correlations for the distribution function Eq.~(\ref{eqn:gpdfdpr}) 
with $V(\varphi) \propto \varphi^{2}$ for the orthogonal ensemble, 
i.e., with the average $S$-matrix $\overline{S}$ been taken as the parameters, 
the correlations are independent of $H_{0}$ while $\overline{S}$ depends on 
$H_{0}$. 
Our purpose of the present article is to analytically show this universality 
in any of the three symmetry classes. 
More precisely, we show 
\begin{equation}
\overline{\prod_{i=1}^{m} \prod_{j=1}^{n} 
\left[ S_{a_{i} b_{i}}\left(E-\frac{\omega}{2}\right) \right]^{k_{i}} 
\left[ S^{*}_{c_{j} d_{j}}\left(E+\frac{\omega}{2}\right)\right]^{l_{j}}} 
= f_{\beta} \bbox{(} \omega \rho(E), \overline{S}(E) \bbox{)}, 
\label{eqn:result}
\end{equation}
where $m$, $n$, $k_{i}$, $l_{j}$ are non negative integers, the bar denotes 
ensemble average. 
The universal functions $f_{\beta}$ depend on the symmetry classes 
($\beta = 1, 2$, and 4 for orthogonal, unitary, and symplectic classes) 
and are independent of $H_{0}$, except for the indices $\{a_{i},b_{i},k_{i},
c_{j},d_{j},l_{j}\}$, while the average local level density $\rho$ 
and the average $S$-matrix $\overline{S}$ depend on $H_{0}$.

\section{The model}

Following the approach of Ref.~\cite{VWZ}, we write the scattering matrix 
$S(E)$ as 
\begin{mathletters}
\begin{equation}
S_{a b}(E) = \delta_{a b} - 2 i \pi \sum_{\mu, \nu} {W^{\dagger}}_{a \mu} 
\left[{D(E)}^{-1}\right]_{\mu \nu} W_{\nu b}, 
\end{equation}
in which 
\begin{equation}
D(E) = E + i 0^{+} - H + i \pi W W^{\dagger}, 
\end{equation}
\end{mathletters}
$E$ is the energy, $0^{+}$ is positive infinitesimal, $H$ represents the 
projection of the full Hamiltonian onto the interaction region, and $W$ 
describes the coupling between the eigenstates of the interaction region and 
the scattering states in the free-propagation region. 
The indices $a$, $b$ refer to the physical scattering channels, 
and $\mu$, $\nu$ refer to the complete orthonormal states 
characterizing the interaction region. 

We assume that $N\times N$ matrix $H$ can be written as 
\begin{equation}
H = H_{0} + \varphi, 
\end{equation}
where $H_{0}$ is a given, nonrandom, Hermitian matrix, and $\varphi$ is 
a member of the Gaussian ensemble.
The symmetry property of $H_{0}$ is the same as $\varphi$.
The independent elements of the matrix $\varphi$ are uncorrelated random 
variables with a Gaussian probability distribution centered at zero. 
The second moments for the unitary ensemble are given by 
\begin{eqnarray}
\overline{ {\varphi}_{\mu \nu}{\varphi}_{\mu' \nu'} } & = 
& \frac{\lambda^{2}}{N} \delta_{\mu \nu'}\delta_{\nu \mu'}. 
\end{eqnarray}
(See Ref.~\cite{M} for the orthogonal and the symplectic cases.) 
Here, $\lambda$ is a strength parameter.

\section{Derivation}

For definiteness, we show the derivation for the unitary ensemble 
and $m, n \le 2$ in Eq.~(\ref{eqn:result}). 
The generalization to the other symmetry classes and/or higher values of 
$m$ and $n$ is straightforward and commented upon in Sec.~\ref{sec:conclu}. 
The derivation is based on the use of Efetov's supersymmetry method 
\cite{VWZ,E}. 
We take the notation from Ref.~\cite{VWZ} and use the [1, 2] block notation 
for the matrix representation in which 1 and 2 refer to the retarded and 
advanced block, respectively.

Consider the following generating function: 
\begin{equation}
Z(J) = 
\frac{ \det \left[ D_{p}(E_{p}) + 2 \pi W J_{p}(F) W^{\dagger} \right] }
     { \det \left[ D_{p}(E_{p}) - 2 \pi W J_{p}(B) W^{\dagger} \right] },
\end{equation}
where $D_{p}(E_{p}) = \mbox{diag}\left[ D(E_{1}), D^{\dagger}(E_{2})\right]$, 
$J_{p}(F) = \mbox{diag}\left[ J_{1}(F), J_{2}(F) \right]$, 
and $J_{p}(B) = \mbox{diag}\left[ J_{1}(B), J_{2}(B) \right]$. 
The scattering matrix can be generated from $Z(J)$ as follows: 
\begin{equation}
{S_{p}(E_{p})}_{a b}
= \delta_{a b}
- i \left. \frac{\partial Z(J)}{\partial {J_{p}(B)}_{b a}} \right|_{J=0} L
= \delta_{a b}
- i \left. \frac{\partial Z(J)}{\partial {J_{p}(F)}_{b a}} \right|_{J=0} L,
\end{equation}
where $S_{p}(E_{p}) = \mbox{diag} \left[ S(E_{1}), S^{\dagger}(E_{2}) \right]$ 
and $L = \mbox{diag}\left( 1, -1 \right)$.
Using standard procedure \cite{VWZ}, we can represent the average of $Z(J)$ 
as an integral over a $4\times 4$ graded matrix field $\sigma$: 
\begin{mathletters}
\begin{equation}
\overline{Z}(J)
=
\int d [\sigma] \, \exp \left\{ {\cal L}(\sigma) \right\}, 
\end{equation}
where
\begin{equation}
{\cal L}(\sigma)
=
- \frac{N}{2\lambda^{2}} \,\mbox{trg} \left( \sigma^{2} \right)
- \mbox{trg} \ln \left[ {\cal D}(\sigma) - \frac{\omega^{-}}{2} L 
+ i \pi W W^{\dagger} L - 2 \pi L_{g} W J_{p}(g) W^{\dagger} \right]. 
\end{equation}
\end{mathletters}
Here, $\mbox{trg}$ denotes the graded trace, 
${\cal D}(\sigma) = E - \sigma - H_{0}$, 
$\omega^{-} = \omega - i 0^{+}$, 
$J_{p}(g)=\mbox{diag} \left[J_{1}(B), J_{1}(F), J_{2}(B), J_{2}(F)\right]$, 
and $L_{g} = \mbox{diag} \left( 1, -1, 1, -1 \right)$. 
We have defined $E = \left(E_{1}+E_{2}\right)/2$ and $\omega = E_{2}-E_{1}$. 

In the limit $N \to \infty$, this integral can be done with the use of the 
saddle-point approximation. 
We are interested in correlations involving energy differences $\omega$ 
of the order of the mean level spacing $\sim O(N^{-1})$. 
Hence, we expand ${\cal L}(\sigma)$ in powers of $\omega$: 
\begin{eqnarray}
{\cal L}(\sigma) & \approx & 
- \frac{N}{2\lambda^{2}} \,\mbox{trg} \left( \sigma^{2} \right)
- \mbox{trg} \ln \left[ {\cal D}(\sigma) \right] 
- \mbox{trg} \ln \left[ 1 + i \pi {{\cal D}(\sigma)}^{-1} 
W W^{\dagger} L \right] \nonumber \\
&&- \mbox{trg} \ln \left\{ 1 - 2 \pi \left[ {\cal D}(\sigma) 
+ i \pi W W^{\dagger} L \right]^{-1} L_{g} W J_{p}(g) W^{\dagger} \right\}
+ \frac{\omega^{-}}{2} \,\mbox{trg} \left[ {{\cal D}(\sigma)}^{-1} L \right]. 
\label{eqn:Lagrangean}
\end{eqnarray}
It should be noted such an expansion is not possible for $W W^{\dagger}$ 
because $W^{\dagger} W \sim O(1)$. 
Of the five terms in expression (\ref{eqn:Lagrangean}) the last three terms 
are $O(1)$. 
The first two terms are $O(N)$ and determine the saddle-point $\sigma^{sp}$. 
To derive the saddle-point equation we write $H_{0}$ and $\sigma^{sp}$ in the 
forms 
$H_{0} 
= U^{-1} \,\mbox{diag} \left( \epsilon_{1}, \ldots, \epsilon_{N} \right) U$ 
and $\sigma^{sp} = T^{-1} \sigma_{D}^{sp} T$, where $\sigma_{D}^{sp}$ is 
diagonal and $T$ has the form 
\begin{equation}
T = 
\left(
\begin{array}{cc}
\left( 1 + t_{12}t_{21} \right)^{1/2} & i \; t_{12} \\
-i \; t_{21} & \left( 1 + t_{21}t_{12} \right)^{1/2}
\end{array}
\right).
\end{equation}
The saddle-point equation reads 
\begin{equation}
\sigma_{D}^{sp} 
= \frac{\lambda^{2}}{N} 
\sum_{\mu =1}^{N} \frac{1}{E - \sigma_{D}^{sp} - \epsilon_{\mu}}.
\label{eqn:spe}
\end{equation}
For ordinary variables (rather than matrices), Eq.~(\ref{eqn:spe}) has the 
$N + 1$ solutions. 
The $N - 1$ of which are real, and the remaining two may have non-zero 
imaginary parts according to the values of $E$. 
Taking the two complex solutions ($r \pm i \Delta$) \cite{BH}, 
we obtain $\sigma_{D}^{sp} = r - i \Delta L$. 
The explicit expressions of $r$ and $\Delta$ are not available
because Eq.~(\ref{eqn:spe}) becomes in general an $(N+1)$-th polynomial. 
Several references discussed the properties of Eq.~(\ref{eqn:spe}) 
(see, e.g., \cite{BHZ,Pastur}). 
Hereafter we consider the case where $\Delta \sim O(1)$. 
From the relation between $\Delta$ and the average level 
density $\rho$ (Eq.~(\ref{eqn:density})), this means 
$E$ lies far away from the edge of the spectrum. 
Substituting $\sigma^{sp}$ for $\sigma$ in Eq.~(\ref{eqn:Lagrangean}), we find 
\begin{eqnarray}
{\cal L}(\sigma^{sp}) & \approx & 
- \mbox{trg} \ln \left[ 1 + i \pi {{\cal D}(\sigma^{sp})}^{-1} 
W W^{\dagger} L \right]
+ \frac{\omega^{-}}{2} 
\,\mbox{trg} \left[ {{\cal D}(\sigma^{sp})}^{-1} L \right] \nonumber \\
&& - \mbox{trg} \ln \left\{ 1 - 2 \pi W^{\dagger} \left[ {\cal D}(\sigma^{sp}) 
+ i \pi W W^{\dagger} L \right]^{-1} W L_{g} J_{p}(g) \right\}.
\label{eqn:SPLag}
\end{eqnarray}

The one-point functions $\rho(E)$ and $\overline{S}(E)$ are evaluated 
at the saddle-point. We thus have 
\begin{mathletters}
\begin{equation}
\rho(E)
= \frac{N \Delta}{\pi \lambda^{2}}
\label{eqn:density}
\end{equation}
and 
\begin{equation}
\overline{S}_{p}(E)
= 1 - 2 i \pi W^{\dagger} \left[ {\cal D}(\sigma_{D}^{sp}) 
+ i \pi W W^{\dagger} L \right]^{-1} W L. 
\end{equation}
\end{mathletters}
Using these one-point functions $\rho(E)$ and $\overline{S}(E)$ 
we can write each term of Eq.~(\ref{eqn:SPLag}) as follows:
\begin{mathletters}
\begin{equation}
\mbox{trg} \ln \left[ 1 + i \pi {{\cal D}(\sigma^{sp})}^{-1} 
W W^{\dagger} L \right]
= \mbox{trg} \ln \left\{ 1 - \left[ \overline{S}_{p}(E) - 1 \right] 
L M \right\}, 
\label{eqn:trgln}
\end{equation}
\begin{equation}
\frac{\omega^{-}}{2} \,\mbox{trg} 
\left[ {{\cal D}(\sigma^{sp})}^{-1} L \right]
=- 2 i \pi \omega^{-} \rho(E) \,\mbox{trg} \left( t_{12} t_{21} \right),
\label{eqn:rho}
\end{equation}
and
\begin{equation}
2 \pi W^{\dagger} \left[ {\cal D}(\sigma^{sp}) 
+ i \pi W W^{\dagger} L \right]^{-1} W
= i \left\{ T^{-1} L \left[ \overline{S}_{p}(E) - 1 \right]^{-1} T 
- T^{-1} M T \right\}^{-1}. 
\label{eqn:inv}
\end{equation}
\end{mathletters}
Here, 
\begin{equation}
M = 
\left(
\begin{array}{cc}
t_{12}t_{21} & - i \; t_{12}\left( 1+t_{21}t_{12}\right)^{1/2} \\
- i \; t_{21}\left( 1+t_{12}t_{21}\right)^{1/2} & - t_{21}t_{12}
\end{array}
\right)
\label{eqn:M}
\end{equation}
and we used the property $T L T^{-1} = L + 2 M$. 
More explicitly, Eqs.~(\ref{eqn:trgln}) and (\ref{eqn:inv}) can be expressed 
with the use of $t_{12}$ and $t_{21}$ as follows: 
\begin{mathletters}
\begin{eqnarray}
\mbox{R.H.S. of Eq.~(\ref{eqn:trgln})} & = & \mbox{trg}\ln \left( 1 
+ {\cal T}_{12} t_{12}t_{21} \right) \nonumber \\
 & = & \mbox{trg}\ln \left( 1 + {\cal T}_{21} t_{12}t_{21} \right) 
\label{TGLE}
\end{eqnarray}
and
\begin{equation}
\mbox{R.H.S. of Eq.~(\ref{eqn:inv})} 
= i \left(
\begin{array}{cc}
\overline{S}(E) \left( 1+ {\cal T}_{21} t_{12}t_{21} \right)^{-1} - 1
& - i\; t_{12} \left( 1+ t_{21}t_{12} \right)^{1/2} 
\left( {{\cal T}_{12}}^{-1} + t_{21}t_{12} \right)^{-1} \\
- i\; t_{21} \left( 1+ t_{12}t_{21} \right)^{1/2} 
\left( {{\cal T}_{21}}^{-1} + t_{12}t_{21} \right)^{-1} 
& 1- \overline{S^{\dagger}}(E) 
\left( 1 + {\cal T}_{12} t_{21}t_{12} \right)^{-1}
\label{INVE}
\end{array}
\right), 
\end{equation}
\end{mathletters}
where ${\cal T}_{1 2} = 1-\overline{S}(E) \overline{S^{\dagger}}(E)$ 
and ${\cal T}_{2 1} = 1-\overline{S^{\dagger}}(E) \overline{S}(E)$ \cite{St}. 
(The derivation is given in Appendix.) 
Thus we find that all the dependence of Eq.~(\ref{eqn:SPLag}) 
on $W$ and $H_{0}$ is completely absorbed in $\rho(E)$ and $\overline{S}(E)$. 
Equations (\ref{eqn:trgln}), (\ref{eqn:rho}), and (\ref{eqn:inv}) show 
universality. 
The explicit forms of correlation functions $f_{\beta}$ are identical with 
those for the Gaussian ensemble and are found in appropriate 
references (see, e.g., \cite{GMW,Be}).

\section{Summary}
\label{sec:conclu}

For the sake of simplicity, we presented the derivation for only the unitary 
ensemble. 
In either of the orthogonal and the symplectic ensemble, the internal 
structures of $t_{12}$ and $t_{21}$ differ from the unitary case. 
However, our derivaton is completely independent of such structures and 
applies equally to the orthogonal and the symplectic ensembles. 
Taking the generating function: 
\begin{equation}
Z(J) = \prod_{ q=1 }^{ \mbox{max} \left\{ m, n \right\} } 
\frac{ \det \left[ D_{p}(E_{p}) + W J_{p}^{q}(F) W^{\dagger} \right] }
     { \det \left[ D_{p}(E_{p}) - W J_{p}^{q}(B) W^{\dagger} \right] },
\end{equation}
we can show Eq.~(\ref{eqn:result}) for $m > 2$ or $n > 2$ along exactly 
parallel lines.

In summary, we have shown that {\it the local universality in the bulk 
scaling limit} still holds for the $S$-matrix correlation functions even 
though unitary invariance is broken by the addition of a deterministic matrix 
to the ensemble. 
The starting random matrix model contains parameters $W$ and $H_{0}$ which 
are specific to individual systems. 
After ensemble averaging, these original parameters are completely absorbed 
into much fewer parameters $\overline{S}(E)$ and $\rho(E)$. 
Thus the $S$-matrix correlaton functions of the type Eq.~(\ref{eqn:result}) 
have universal forms which are independent of $H_{0}$ but for 
$\overline{S}(E)$ and $\rho(E)$ and are determined only by the symmetry of the 
ensemble. 
This holds for all the three symmetry classes (orthogonal, unitary, and 
symplectic). 
The derivation can be similarly applied to the spectral correlation functions. 
Thus we have extended the previous results by Br\'ezin and Hikami \cite{BH} to 
the orthogonal and the symplectic ensembles though only two-point functions 
are considered. 

The present results were derived under the restrictions that the correlation 
functions contain only two values of energy, $E_{1}$ and $E_{2}$, and that 
$V(\varphi)$ has a Gaussian form. 
It is a natural conjecture that the universality of the $S$-matrix correlation 
functions holds even if these two restrictions are removed. 
The increase of the number of energy arguments makes the structure of 
$\sigma_{D}^{sp}$ more complicated. 
With this point taken properly into account, the similar derivation is 
probable. 
The extension to the general form of $V(\varphi)$ seems less trivial because we are no 
longer able to use a Hubbard-Stratonovich transformation in order to introduce 
a graded matrix $\sigma$. 
The simlar procedure used in Ref.~\cite{HW} may be incorporated into present 
derivation. 

\section*{Acknowledgments}

Stimulating discussions with K. Nohara and K. Takahashi are appreciated with 
thanks.

\appendix

\section*{Derivation of Eqs.~(\ref{TGLE}) and (\ref{INVE})}

To derive Eqs.~(\ref{TGLE}) and (\ref{INVE}), we use the fact that 
for any analytic function $F$, we have 
$t_{12} F(t_{21} t_{12}) = F(t_{12} t_{21}) t_{12}$. 
Using the identity
\begin{equation}
\mbox{trg} \ln
\left(
\begin{array}{cc}
a & b \\
c & d
\end{array}
\right)
= \mbox{trg} \ln \left( a - b\, d^{-1} c \right) 
+ \mbox{trg} \ln \left( d\, \right),
\end{equation}
we obtain Eq.~(\ref{TGLE}).
For the abbreviation $Y \equiv \left\{ T^{-1} L \left[ \overline{S}_{p}(E) 
- 1 \right]^{-1} T -  T^{-1} M T \right\}^{-1}$, using the property 
$T^{-1} = L T L$, we get the following equation 
\begin{equation}
Y = 
\left(
\begin{array}{cc}
\left[ \overline{S}(E) - 1 \right]^{-1} + t_{12} t_{21} A 
& i \; t_{12} \left( 1 + t_{21}t_{12} \right)^{1/2} A \\
i \; t_{21} \left( 1 + t_{12}t_{21} \right)^{1/2} A
& - \left[ \overline{S^{\dagger}}(E) - 1 \right]^{-1} - t_{21} t_{12} A
\end{array}
\right)^{-1}, 
\end{equation}
where $A = \left[ \overline{S}(E) - 1 \right]^{-1} 
+ \left[ \overline{S^{\dagger}}(E) - 1 \right]^{-1} + 1$. 
With the use of the formula
\begin{equation}
\left(
\begin{array}{cc}
a & b \\
c & d
\end{array}
\right)^{-1}
=
\left(
\begin{array}{cc}
\left( a - b\, d^{-1} c \right)^{-1}
& - a^{-1} b \left( d - c\, a^{-1} b \right)^{-1} \\
- d^{-1} c \left( a - b\, d^{-1} c \right)^{-1}
& \left( d - c\, a^{-1} b \right)^{-1}
\end{array}
\right),
\end{equation}
the [1, 1] block of $Y$ can be written as follows 
\begin{equation}
Y_{11} = A^{-1} \left\{ \left[ \overline{S^{\dagger}}(E) - 1 \right]^{-1} 
+ t_{12} t_{21} A \right\} 
\left\{ \left[ \overline{S}(E) - 1 \right]^{-1} A^{-1} 
\left[ \overline{S^{\dagger}}(E) - 1 \right]^{-1} 
- t_{12} t_{21} \right\}^{-1}.
\label{eqn:inmi11}
\end{equation}
Using the property 
\begin{eqnarray}
A & = & \left[ \overline{S^{\dagger}}(E) - 1 \right]^{-1} 
\left[ \overline{S^{\dagger}}(E) \overline{S}(E) -1 \right] 
\left[ \overline{S}(E) - 1 \right]^{-1} \nonumber \\
  & = & \left[ \overline{S}(E) - 1 \right]^{-1} 
\left[ \overline{S}(E) \overline{S^{\dagger}}(E) -1 \right] 
\left[ \overline{S^{\dagger}}(E) - 1 \right]^{-1}, 
\end{eqnarray}
we obtain the [1, 1] block of Eq.~(\ref{INVE}). 
Similarly the [2, 1] block of $Y$ can be written as follows 
\begin{equation}
Y_{21} = i \; t_{21}\left( 1+t_{12}t_{21} \right)^{1/2} 
\left\{ \left[ \overline{S^{\dagger}}(E) - 1 \right]^{-1} 
+ t_{12} t_{21} A \right\}^{-1} A\, Y_{11}.
\label{eqn:inmi21}
\end{equation}
Substituting Eq.~(\ref{eqn:inmi11}) for $Y_{11}$ in Eq.~(\ref{eqn:inmi21}), 
we obtain the [2, 1] block of Eq.~(\ref{INVE}). 
The other blocks are obtained along exactly parallel lines.


\begin{references}
\bibitem{St} The transmission coefficients (or sticking probabilities) in 
Ref.~\cite{VWZ} correspond to the eigenvalues of ${\cal T}_{12}$ (which are 
equal to those of ${\cal T}_{21}$).
\end{references}
\end{document}